\documentclass[12pt,preprint, a4paper, onecolumn]{aastex}

\usepackage{amsmath}
\usepackage{epsfig}

\newcommand{\be}{\begin{equation}}
\newcommand{\cm}{{~\:\rm cm}}

\newcommand{\ee}{\end{equation}}
\newcommand{\erg}{{\rm erg}}
\newcommand{\ergs}{{\rm erg\; s^{-1}}}

\newcommand{\gcc}{{\rm g \; cm^{-3}}}
\newcommand{\K}{{~\:\rm K}}
\newcommand{\keV}{{~\:\rm keV}}

\newcommand{\kms}{{~\:\rm km\, s^{-1}}}
\newcommand{\kpc}{{~\:\rm kpc}}

\newcommand{\msyr}{{~\: M_{\odot}\;\rm yr^{-1}}}
\newcommand{\pc}{{~\:\rm pc}}
\newcommand{\yr}{{~\:\rm yr}}

\begin{document}



\title{Feedback Heating with  Slow Jets in  Cooling Flow Clusters}


\author{Noam Soker}
\affil{Department of Physics, Technion--Israel Institute of Technology, 
Haifa 32000}
\email{soker@physics.technion.ac.il}

\and

\author{Fabio Pizzolato}
\affil{Department of Physics, Technion--Israel Institute of Technology, 
Haifa 32000}
\email{fabio@physics.technion.ac.il}


\begin{abstract}
We propose a scenario in which a large fraction, or even most, of
the gas cooling to low temperatures of $T<10^4 \K$ in cooling flow
clusters, directly gains energy from the central black hole.
Most of the cool gas is accelerated to non-relativistic high
velocities, $v_j \simeq 10^3- 10^4 \kms$, after flowing
through, or close to, an accretion disk around the central black hole.
A poorly collimated wind (or double not-well collimated
opposite jets) is formed.
According to the proposed scenario, this gas inflates  some of the
X-ray deficient bubbles, such that the average gas temperature
inside these bubbles (cavities) in cooling flow clusters is
$kT_b \lesssim 100 \keV$.
A large fraction of these bubbles will be very faint, or
not detectable, in the radio.
The bright rims of these weak smaller bubbles will appear as ripples.
We suggest that the X-ray ripples observed in the Perseus cluster,
for example, are not sound waves, but rather the rims of
radio-faint weak bubbles which are only slightly hotter
than their environment.
This scenario is incorporated into the moderate cooling flow
model; although not a necessary ingredient in that model,
it brings it to better agreement with observations.
In the moderate cooling flow model a cooling flow does
exist, but the mass cooling rate is $\lesssim 10\%$ of
that in old versions of cooling flow models.
\end{abstract}


\section{Introduction}
\label{s:intro}

Recent {\sl Chandra} and {\sl XMM-Newton} observations of cooling
flow (CF) clusters of galaxies have failed to detect the large
amounts of cool gas predicted by the old versions of the cooling
flow (CF) model \citep[e.g.][]{Tam01, Pet03, Pet04, Mol01}, 
reducing by at least  one order
of magnitude the expected mass cooling rate to low temperatures.
These results raise several questions. Do CFs occur at all? If
they do, what is the actual mass cooling rate? What is the fate of
the cool gas? In many clusters an unhindered  radiative loss would
lead to a sizeable cooling flow in few $10^8$~yr:  since this does
not seem to happen, some reheating is required \citep[e.g.][]{Fab03b},
what is the nature of this heating mechanism?

In this paper we adopt the framework of the moderate CF model
to address the fate of the cool gas, and suggest  that a
large fraction, and in some cases most, of the cool  gas is being
ejected back to the ICM by the  active galactic nucleus (AGN)
operating at the central cD galaxy. The approach  we adopt here
differs from those adopted by most authors 
(e.g. 
\citeauthor{Bin95}  \citeyear{Bin95};
\citeauthor{Tuc97}  \citeyear{Tuc97};
\citeauthor{Cio01}  \citeyear{Cio01};
\citeauthor{Bin04}  \citeyear{Bin04};
\citeauthor{Omm04a} \citeyear{Omm04a};
\citeauthor{Mat04}  \citeyear{Mat04};
for more references see \citeauthor{Pet04}  \citeyear{Pet04}),
in that we {\em do} consider
a CF, moderate one though. We shall conclude that the best
agreement with the available data occurs if the jets are not well
collimated.

In \S~\ref{s:mcf}  we shall sketch the envisaged scenario, putting it
in the appropriate framework of the moderate CF model.
In \S~\ref{s:hot} we review some hints to slow jets. Among them, a
key role is played by the hot bubbles observed with {\sl Chandra}
in several CF clusters.
In \S~\ref{s:jet} we shall discuss the properties of slow and massive
jets, and learn that the   bubbles' properties seem to require slow
and dense jets, which imply mass cooling rate of
$\gtrsim 10 \msyr$ in large CF clusters.
In other words,  {\em the bubbles imply the presence of a CF.}
We summarize in \S~5.

\section{The Proposed  Scenario}
\label{s:mcf}

We start this Section by recalling some important features of the
moderate CF model  \citep[for more details, see][]{Sok01, Sok03}, 
as well as some other relevant  ingredients for  our proposed model.

The moderate CF model, which was proposed before the new results from
{\sl Chandra} and {\sl XMM-Newton} (hence we avoid referring to 
old version of
the CF model as {\em standard}), is different from many earlier proposed
processes whose aim is to prevent CF in clusters of galaxies altogether
\citep[][]{Sok01,  Sok03}.
The main ingredient of the moderate CF model is that the effective age,
i.e. the time elapsed since the last major disturbance of the
intra-cluster medium (ICM) inside the cooling radius
$r_c \simeq100 \kpc$, is much shorter than the cluster age
\citep[e.g.][]{Bin95, Bin04, Sok01}.
The cooling radius is defined as the place where the radiative
cooling time equals the cluster age.
Originally, the heating in the moderate CF model was proposed to be
intermittent \citep[][]{Sok01}, but the basic idea can hold
for a steady heating, or heating in short intervals 
\citep[][]{Bin04}.

In the moderate CF model 
most of the gas within the CF resides in the hottest phase, which is
prevented from cooling continuously and attaining a steady-state
configuration by being reheated \citep[][]{Sok03,Kai04}.
This results in a mass cooling rate that decreases with decreasing
temperature, with a much lower mass cooling rate at the lowest
temperatures.
The limit on the cooling rate below a temperature $T_{\rm min}$
inferred from X-ray observations is $< 20 \%$ of the
mass cooling rates cited in the past 
\citep[e.g.][]{Fab02, Mol01, Pet03}.
In some cases, though, cooling to low temperatures is observed,
as in the CF clusters Abell~2597 \citep[][]{Mor04}
and Abell~2029 \citep[][]{Cla04}.
In Abell~2597 both extreme-UV and X-ray observations indicate a mass
cooling rate of $\sim 50 \msyr$, which is $\sim 0.2$
of the value quoted in the past based on {\sl ROSAT} X-ray observations
\citep[see the discussion in][]{Mor04}.
Some fraction of the gas may cool to lower temperatures by heat
transfer to optical-emitting gas, reducing further the
X-ray emission from gas residing at temperatures of
$T \lesssim T_{\rm min} \simeq 1 \keV$;
the energy transfer can be via mixing
\citep[e.g.][]{Oeg01, Fab01, Fab02, Joh02, Bay02},
and/or heat conduction \citep[][]{Sok04c, Sok04b}.
Intermittent heating, which is part of the moderate CF model,
probably occurs by AGN activity.
In a recent paper \citet{McN04a} found a huge deficient
X-ray bubble pair in the cluster MS~0735.6+7421, which suggests
that an intermittent heating over a large time scale might
play a role.
In the moderate cluster CF model, the agreement between
star formation rate and mass cooling rate can be quite good.
\citet{Wis04} and \citet{McN04b},
for example, find the cooling rate within $r \simeq30 \kpc$ of the CF
cluster A~1068 to be about equal to the star formation rate there
($20-70\msyr$).

Another plausible ingredient of the moderate CF model is that the
entire inner CF region supplies the cold gas
accreted to the black hole \citep[][]{Piz04}.
Namely, the feedback between heating and cooling
occurs with the entire cool inner region,
$r \lesssim 5-30 \kpc$, in what we term a {\em cold-feedback model}.
In the proposed scenario \citep[][]{Piz04} non-linear
over-dense blobs of gas, $\delta \rho /\rho_a \gtrsim 2$,
i.e., $\rho/\rho_a \gtrsim 3$, in this inner region cool on
short time scales such that they are removed from the
ICM before the next major AGN heating event in their region.
Some of these blobs cool and sink toward the central black hole,
while others may form stars and cold molecular clouds.
This mechanism can work on the condition that the blobs have a small angular
momentum, or they  would be   prevented from accreting onto the central black 
hole and fuelling its activity.
This issue has been discussed  in a separate paper \citep[][]{Piz04}: here
we just summarize the main conclusions. The cold blobs 
may stem  directly from  ICM disturbances  driven by an earlier AGN activity,
but also  from galaxies mass-stripping \citep[][]{Sok91}.
Since the  galaxies do not have an ordered bulk motion, the mass
stripped from them is also unlikely to conspire and organize in an
ordered  flow with high net angular momentum.
Therefore, if a circular flow like a disc  forms,
it cannot be very large: say, $\sim 10^2 \pc$ as in M87
\citep[][]{Har94,For94}. 
The orbit of a  dense blob subjected to gravity and the friction drag force 
has  a circularization radius whose size  is  in broad agreement 
with this value \citep[][]{Piz04}.
Besides,  in the cold  feedback model the cold gas
is expected to accrete from regions not too far from the center (few tens kpc
at  most), so the accreting flow should  not have a large angular 
momentum from  the outset. Therefore,   the blobs' angular momentum 
is not high enough to prevent them from accreting in a time scale
shorter than or comparable to the  cooling time, which keeps running the  
cooling/accretion feedback loop.

In the present paper we suggest that a large fraction, and in some cases most,
of the accreting cool gas is being ejected back to the ICM by the AGN.
This idea solves some problems related to the questions posed above.
In particular, the slower and more massive collimated wind
(namely, a poorly collimated jet pair; hereafter referred to
simply as jets), deposits its energy in the inner region
\citep[][]{Bin04}, as required by recent observations that limit the
degree of mixing \citep[][]{Boh04}, and can carry more
energy than inferred from radio observations \citep[][]{Bin04}.
In the present paper we extend this idea of slow and massive
not-well collimated jets, and incorporate it to be an important
ingredient in the duty-cycle of the moderate CF model.
The proposed scenario is fundamentally different from the scenario of
\citet{Nul04} or that of \citet{Bin04} \citep[see also][]{Omm04b}.
In those papers, most, or all, of the mass which is accreted to
the central black hole comes from the hot phase, $T \simeq 10^7 \K$.
Hence, the mass that is ejected in the jets is small compared to
the mass assumed to be cooling to $T \lesssim 10^4 \K$ in the
moderate CF model.
In the proposed scenario, on the other hand, we address the
question of the fate of the cold gas, arguing that non negligible
fraction of it is ejected back to the ICM at high non-relativistic
velocities.
In those papers the duty-cycle, or feedback process, is determined
by energy considerations alone, while in the present scenario
the mass is also a factor in the duty-cycle.

Our proposed scenario is also different from the circulation flow proposed
by \citet{Mat04}; in their scenario the gas does not cool below
X-ray emission temperature.
This is a significant difference, as basically they don't consider the
presence of a CF to low temperatures ($T< 10^4 \K$),
while we do. 
In common with their model, we ascribe significance to mass as well
as energy transport in both inward and outward direction.

The cool ICM flows to the center, and if it has  an angular momentum, 
it may form an accretion disc about the central black
hole. A  standard geometrically  thin, optically thick accretion discs
(\`a la Shakura-Sunyaev) truncated at the last marginally stable
orbit  around the black hole  has a radiative  efficiency $\eta\simeq 10\%$,
corresponding to a bolometric luminosity
\begin{equation}
\label{e:lacc}
L_{\rm accr} \simeq 5.7 \times 10^{45} \ergs \;
\left(\frac{\eta}{0.1}\right)
\left( \frac{\dot{M}}{\msyr} \right),
\end{equation}
Such bright discs are not common, which forces us to assume that the disc's
radiative efficiency is low. The disc may be truncated at several
Schwarzschild radii from the hole (e.g. it may be evaporated  by
magnetic fields), which reduces $\eta$ in Equation~\eqref{e:lacc}
by orders of magnitude. 
Alternatively, the disc may be radiatively inefficient, i.e. it does not
radiate most of the  energy converted by  the viscous dissipation: such is
an {\sl ADAF} or an {\sl ADIOS} \citep[][]{Nar94, Bla99}.
\citet{Omm04a} assume an {\sl ADIOS} flow in the
framework of  their simulations.
In these kinds of accretion flow,  most of the incoming flux is
gravitationally unbound to the black hole, and can therefore be pushed back
as an outflow, or a wind.
A possible issue it  that  {\sl ADAF}s cannot exist for high accretion rates
\citep[][]{Qua99}. There is the possibility, however,
that the disc is composite, with the {\sl ADAF} component confined in the 
inner part, the outer being an ordinary accretion disc 
\citep[e.g.][]{Qua99}.
Discs like these may exist even for
high accretion rates, and may also support the required outflow.

In the next section we start by reviewing some hints that the outflowing  
jets are slow. It seems as if the properties of bubbles require slow
and dense jets, which imply mass cooling rate of
$\gtrsim 10 \msyr$ in large CF clusters.
Namely, {\em  the bubbles imply the presence of a CF.}
We then estimate the relevant parameters, and summarize by listing
some predictions of the proposed model.

\section{The Temperature of Hot Bubbles}
\label{s:hot}

We list below some pieces of evidence that some bubbles are inflated 
mainly by non-relativistic jets, i.e.,  the temperature of the gas inside
some bubbles is $T_b \lesssim 100 \keV$.  A population of
radio emitting relativistic electrons may exist due to the contribution from
a relativistic jet as well.
In other cases, like in the large bubbles in M87, the inflating jets are
relativistic \citep[][]{For03}, and the non-relativistic component is lacking.
\begin{enumerate} 
\item
In most clusters  only  a  lower limit determination 
of the hot bubbles temperature is possible \citep[][]{Bla04}. 
Some examples are 
$T_b > 15 \keV$ for Hydra~A \citep[][]{Nul02},
$T_b > 11 \keV$ for Perseus \citep[][]{Sch02}, and
$T_b > 20 \keV$ for Abell~2052 \citep[][]{Bla03}.
We are aware  only of  the case of MKW~3S 
where a bubble  temperature  was  measured \citep[][]{Maz02}.
In this cluster a surface brightness depression located at a distance of
$r_b \simeq 100 \kpc$ from the cluster center has a 
temperature  $T_b \simeq 6-10 \keV$ \citep[][]{Maz02}.
Both in Hydra~A \citep[][]{Dav01} and Abell~2052 \citep[][]{Bla01}
the pressure drops by a factor of $f_p \simeq 0.1$ from the inner region to
$r \simeq 100 \kpc$.
For an adiabatic expansion with a constant ratio of the bubble to
external pressure, the temperature inside the bubble drops by
a factor of $f_t =f_p^{2/5} \simeq 0.4$.
If we assume that the bubble in MKW~3S was inflated in the
inner region, and then was buoyantly rising and
adiabatically expanding, then its initial temperature was
$T_{b0} \simeq 15-25 \keV$.
The preliminary results of \citet{McN04a} show that the large
two X-ray deficient bubbles in the cluster MS~0735.6+7421 are only slightly
hotter than their environment, strengthening the results above.
A word of caution about these measurements is in order, however. 
It is extremely difficult to 
measure the  temperature of the material in the ICM bubble with the 
current data. The  geometry of the cavities is never known well 
enough to allow an accurate   deprojection;  besides  it is 
unknown whether the emission stems  from the body of the cavity or from its
rims.  If we were to adopt a more conservative viewpoint, we could say
that the current temperature measurements are  {\em consistent} with 
the view  that these blobs are filled with hot plasma.
\item The pressure in the bubbles inferred from the radio emission
and the assumption of equipartition between magnetic and relativistic
particles is lower than the ambient pressure \citep[][]{Bla04}.
This hints (but does not {\em prove}),  that the bubbles are not 
entirely  relativistic, because the bubbles are in pressure equilibrium with
their environment.
\item
Some AGNs have been observed to blow non-relativistic winds, with speeds 
down to  $24,000 \kms$ \citep[PDS~456:][PG1211+143: \citeauthor{Pou03} 
\citeyear{Pou03}; however, see \citeauthor{McK04} \citeyear{McK04} for some
possible problems]{Ree03}. 
The identification of an even slower wind
($\sim 1,000 \kms$) in the {\sl LINER} NGC~1097 is more problematic 
\citep[][and  references therein]{Sto03}. 

On the theoretical side, \citet{Bin04} lists some arguments
in support of a slow outflow.  According to \citet{Bin04},
in many cases the relativistic jet carries a small fraction of the
mass and energy in the outflow, and in some systems no
relativistic jet is present even when slower outflow occurs.
In their bubble-inflation simulations, \citet{Omm04b}
take the wind speed to be in the range $2.8-4 \times 10^4 \kms$,
with a mass loss rate of $1 \msyr$,
and \citet{Omm04a} take $v_j=10,000 \kms$ and
a mass loss rate into the jet of $2 \msyr$.
In many models for quasars the velocity of the wind blown by the disk
is in the range $\sim 10^3 - 3 \times 10^4 \kms$
\citep[][and references therein]{Elv00, Cre03}.
There are other models which predict slow ($\lesssim 10^4 \kms$)
winds, with the most recent one being the one by \citet{Beg04}.
In the model proposed by \citet{Nic00}, the wind velocity decreases
from $\sim 20,000 \kms$ to $\sim 1000 \kms$ as
the mass accretion
rate increases; in this model the high accretion rate expected in
CF clusters may result in slow winds.
In Seyfert galaxies, mass loss rates from the black hole vicinity as large
as $\simeq 1\msyr $ have been inferred \citep[][]{Cre03}.
The outflow of the absorbing gas starts at distances as small as
$\sim 0.01\pc$ from the black hole, indicating origin from the
accretion disk.
The possibility of slow jets from the nucleus (3C~317) of the cooling
flow cluster A2052 was raised recently by \citet{Ven04}.
\end{enumerate}

The slow wind velocities suggested above leads to bubble temperatures
of $kT_B \lesssim 100 \keV$.
Using the expression given by \citet{Cas75} for the
expansion of an interstellar bubble
\citep[][also applied this expression to study the
bubble formed by the radio jet in M87]{Bic96}, the temperature
inside the bubble is
\begin{equation}
\label{e:temp}
kT_B \simeq 0.15 \mu m_H v_j^2
= 100 \left( \frac{v_j}{10^4 \kms} \right)^2 \keV,
\end{equation}
where $v_j$ is the jet velocity, and $\mu m_H$ the mean mass per particle.
This temperature implies a low density in the bubble, assuming
pressure equilibrium with the ICM.
Even for $v_j =4000 \kms$, the bubble temperature is
$\gtrsim 5$ times higher than the ambient temperature in the
inner regions of CF clusters (typically $\simeq1-3 \keV$).
This means a bubble density of $n \lesssim 0.2$ times the ambient
density, and emissivity $n^2 \lesssim 0.04$ times that of the ambient
medium.
This low emissivity cannot be detected by present X-ray telescopes.

If, as assumed here following \citet{Bin04}, the AGN blows gas
with a spectrum of velocities, then the bubble is multi-phase.
In some cases the fastest blown gas has relativistic speeds,
and the bubble is detectable in the radio.
The formation process of the bubble is likely to lead to some internal
motion within the bubble.
This internal motion determines the mixing of the different
gas-phases at later times, and the magnetic field topology
and evolution determines local heat conduction.
The evolution at late times is hard to predict,
and probably very hard to simulate as well.
We do expect that the phases will be mixed among themselves,
and later on with the ICM, hence heating the ICM.
As with other bubbles models, the bubbles themselves heat
the ICM by doing work on the ICM, and by lifting cool
ICM medium from the center.

We stress  that in the present  model  slow winds do {\em not} 
exclude   the occurrence of fast, relativistic jets.
Indeed, many of the best examples of ICM cavities are clearly associated 
with radio lobes 
(e.g. 
Hydra-A:  \citeauthor{McN00}  \citeyear{McN00};
Perseus:  \citeauthor{Fab02a} \citeyear{Fab02a}).
Slow winds and relativistic jets may coexist:  
the former carry most of the energy  \citep[][]{Bin04},
and the latter
fuel of the synchrotron emission from the  relativistic electrons in the 
X-ray cavities.

\section{Jet Properties}
\label{s:jet}

\subsection{The Mass Flow Rate into the Jets}
\label{s:jet:mdot}

Combining the estimated energy of bubbles (or bubble pairs)
with the typical non-relativistic velocity assumed in the
previous Section  yields the mass ejection rate.
\citet{Bir04} give the observed
mechanical luminosity of 16 clusters.
The jets' kinetic luminosity will be equal to several
times the power $L_{\rm mech}$ calculated by them from the
bubbles values of $PV$, where $P$ is the pressure and $V$
the bubbles' volume.
Most luminous clusters reside, therefore, in the range
$\dot E_j \simeq 0.3-3 \times 10^{44} \ergs$, with
possible higher values if the pressure inside bubbles
is higher than in their surrounding \citep[][]{Bir04},
or if some energy is in shocks formed by the inflated bubbles
\citep[][]{For03, McN04a}.
Numerical simulation require the power to be on the upper side
$\dot E_j \simeq5 \times 10^{44}\ergs$
\citep[e.g.][]{DVe04, Omm04b}.
Also, the power of jets in AGN can reach a power of
$\dot E_{j,\; {\rm max}} \simeq10^{47}\ergs$ \citep[e.g.][]{Raw91}.
We therefore scale the averaged energy injected by
the AGN in CF clusters with values higher than the
energy directly observed \citep[][]{Bin04}, and take
it to be $5 \times 10^{44}\ergs$.
Scaling with typical values and using Equation~\eqref{e:temp}, the mass
loss rate into the jets is
\begin{equation}
\label{e:mjet}
\dot M_j \simeq 15
\left( \frac {\dot E_j} {5\times 10^{44}\ergs} \right)
\left( \frac {kT_b} {100 \keV} \right)^{-1} \msyr.
\end{equation}
In MKW~3S the temperature inside the bubble $T_b$ has been estimated
\citep[][]{Maz02}, and we take the value calculated in the
previous section for the initial value of the temperature,
$T_{b0} \simeq 20 \keV$, and take also
$\dot E \gtrsim 10^{44}\ergs$ found by \citet{Bir04}.
From these we find $\dot M_j \gtrsim 15 \msyr$.
This is more than the X-ray-inferred mass cooling rate of
$\dot M_{\rm cool} < 2 \msyr$ \citet{Bir04}.
However, in MKW~3S the bubbles are not so prominent as in
other clusters, and may represent a relatively old
ejection event.

\subsection{Jet Propagation}
\label{s:jet:prop}

Based on \S~3 of \citet{Sok04a}, we note the following properties
in regards to the heavy jets.
It is assumed that the slow collimated wind (or not-well collimated jet)
has a wide opening angle, measured from the symmetry axis of the jet,
of $\alpha \simeq1$.
The expansion velocity of the jet's head $v_h$ is given by the following
expression  as long as $v_h \ll v_j$:
\begin{eqnarray}
\label{e:vhead}
v_h \simeq \left[ \frac {\dot E_j}{\pi (1-\cos \alpha) z^2 v_j \rho_c} \right]^{1/2}
= 1200  
\left( \frac {\dot E_j} {5 \times 10^{44}\ergs} \right)^{1/2}
\left( \frac {v_j}{10^4 \kms} \right)^{-1/2}
\nonumber \\ \times
\left( \frac{1-\cos \alpha}{0.5} \right) ^{-1/2}
\left(\frac {\rho_c}{10^{-25} \gcc} \right)^{-1/2}
\left( \frac {z} {5 \kpc} \right) ^{-1} \kms,
\end{eqnarray}
where $\rho_c$ is the ambient density, and the $z$ the distance of the
jet's head from its source, measured along the jet's symmetry axis.
Note that the jet is scaled with an opening angle of $\alpha = 60^\circ$
(from its symmetry axis; the full opening angle is $120^\circ$).
For these parameters the jet becomes subsonic in the ICM at
a distance of $\sim 10 \kpc$.
The subsonic jet expands to the side, ensuring the formation
of a large bubble.
A large opening angle, although not a necessary condition to
inflate a bubble, facilitates the formation of a large
bubble close to the center.

A bubble can be formed before the jet's head becomes subsonic if
the shocked material in the jet expands faster than the jet's head
\citep[][]{Sok04a}.
The condition on the opening angle of the jet for that to occur
is given by Equation~(14) of \citet{Sok04a}.
We change the variable in that equation as follows.
From Equation~\eqref{e:vhead}, the distance of the jet's head as function of
time is given by
\begin{eqnarray}
z \simeq 3.4  
\left( \frac {t} {10^6 \yr} \right) ^{1/2} \kpc.
\end{eqnarray}
We substitute the value of the time from the last equation into
Equation~(14) of  \citet{Sok04a} to derive the condition to inflate
a bubble via the mechanism discussed in \citet{Sok04a}
\begin{eqnarray}
\label{e:alpha}
\alpha \gtrsim 50 ^\circ
\left( \frac {\dot E_j} {5 \times 10^{44}\ergs} \right)^{3/10}
\left( \frac {v_j}{10^4 \kms} \right)^{-1/2}
\left(\frac {\rho_c}{10^{-25}\gcc} \right)^{-3/10}
\left( \frac {z} {10 \kpc} \right) ^{-1/5}.
\end{eqnarray}

From Equations~\eqref{e:vhead} and~\eqref{e:alpha} we see that the large
opening angle
of the jet (hence termed here a not-well collimated jet) facilitates
bubble formation  \citep[][]{Sok04a}.
In the proposed scenario a bubble is formed at a distance of
$z \simeq10 \kpc$ from the source of the jet.
The large opening angle assumed here is different from the
heavy slow jet simulation of \citet{Omm04a}.

The large opening angle implies that the jet, when expands outward,
interacts with a substantial fraction of the ICM in its vicinity.
The jet, and the bubble it forms, pushes outward
ICM medium which is cooler and denser than the ICM the bubble
reaches at later time.
Such a wide open angle flow might account for dense shells observed
around some bubbles, e.g., in Abell~2052 \citep[][]{Bla01}.

With our proposed scenario, where many small and weak bubbles which
are only slightly hotter than their environment exist, we turn to the
deep X-ray image of the Perseus CF cluster \citep[][]{Fab03c}.
Many bright arcs, which are termed ripples, are seen in this cluster.
They were interpreted by \citet{Fab03c} as sound waves.
However, we suggest that these are actually the bright rims of 
a stack of weak flattened bubbles accumulated in the past, which 
have been flattened by  the resistance of the environment ICM they
push through.
This is supported by noting the following:
(1) The rims of the two inner X-ray deficient strong bubbles, and
the rim of the outer strong bubble, look like the X-ray ripples, both
in shape and size; only that these rims are brighter than the ripples.
(2) Some ripples, when considered  as geometric spherical arcs, have their
center off from the cluster center.
(3) While sound waves are expected to expand to all directions,
some ripples are very short.
One possible prediction of our proposal that the ripples are actually
rims of weak bubbles, is that very weak radio emission will be detected
between some ripples, similar to, but much weaker than, the radio emission
in X-ray deficient bubbles.

On the other extreme, a slow jet can be well collimated, hence
propagates along a narrow cone into the ICM.
Let $\beta$ be defined such that the jet expands into
a solid angle of $\Omega_j = 4 \pi \beta$.
For $\alpha=12^\circ$, for example, $\beta=0.01$.
By neglecting the magnetic pressure inside the jet and relativistic
effects, hence $\dot E_j = \dot M_j v_j^2/2$, the speed of the
jet's head is determined by pressure equilibrium on its two sides.
An approximate relation is obtained if we consider only ram pressures,
$\rho_j(v_j-v_h)^2=\rho_a v_h^2$, where $\rho_a$ is the ICM density,
$\rho_j = \dot M_j/(4 \pi \beta r^2 v_j)$,
is the density inside the jet, $v_j$ is the speed of the gas inside
the jet, $v_h$ the speed of the jet's head, $\dot M_j$ is the mass
loss rate into one jet, and $r$ is the distance of the jet's head
from its origin \citep[e.g.][]{Kra03}.
For $v_j \gg v_h$, the jet's head propagation speed is given by
\begin{eqnarray}
\label{e:v2}
v_h =\frac{v_j}{(\rho_a/\rho_j)^{1/2} +1} \simeq 600   
\left( \frac {\dot E_j} {5 \times 10^{43}\ergs} \right)^{1/2}
\left( \frac {v_j}{10^5 \kms} \right)^{-1/2}
\nonumber \\ \times
\left( \frac{\beta}{0.01} \right) ^{-1/2}
\left(\frac {n_a}{0.1 \cm^{-3}} \right)^{-1/2}
\left( \frac {r} {5 \kpc} \right) ^{-1} \kms.
\end{eqnarray}
We scaled the power of one jet according to that of Hydra~A as given by
\citet{Bir04}, the total ambient number density is scaled with
that of Hydra~A as given by \citet{Dav01},
and we scaled with a very fast jet of $v_j=0.33c$.

The last equation shows that jets like those in Hydra~A, if they are
very fast, become subsonic at $\sim 5 \kpc$ from their origin.
This explains the observations that the jets in Hydra~A turn into
lobes.
But what is the reason for the sharp transition from a well collimated
jet to radio lobes in Hydra~A?
We note from the last equation that a slow jet, even if weak,
might stay supersonic to large distances.
For example, for a jet with $\dot E = 10^{43} \erg s^{-1}$ and
$v_j=3000 \kms$ we find $(\rho_a/\rho_j)^{1/2} \simeq 2$, and
the exact solution of the last equation is $v_h \simeq 1000 \kms$,
at a distance of $r=5 \kpc$.
We propose, therefore, that the current jets in Hydra~A were preceded
by collimated slow and dense jets which opened a tunnel to a distance of
$r \sim 5 \kpc$, through which the current jets are expanding almost
undisturbed.
When the current jets leave these tunnel, they interact with the ICM,
become subsonic, and lose their collimation.
A slow jet with these parameters ($\dot E = 10^{43} \erg s^{-1}$
and $v_j=3000 \kms$) has  a mass loss rate of
$3.5 \msyr$, namely, the two proposed slow jets blow
$\sim 7 \msyr$ back to the ICM.
To propagate to a distance of $\sim 5 \kpc$, the slow jets were
active for few$\times 10^6 \yr$.
Hence, in total, they blew a non negligible mass back to the ICM.

\subsection{Spectral signature of the outflow}

What are the observational characteristics of the slow poorly collimated
outflow?  Usually, the winds blown by  AGNs are revealed by their 
absorption of the underlying continuum  in the UV or X-rays
(e.g. 
PG~1211+143: \citeauthor{Pou03} \citeyear{Pou03};
PDS~456:     \citeauthor{Ree03} \citeyear{Ree03};
NGC~1097:    \citeauthor{Sto03} \citeyear{Sto03};
see also     \citeauthor{Cre03} \citeyear{Cre03} for a recent review).
Since the  outflows we have dealt with in this paper are not very hot at 
their outset ($10^5-10^6\K$), also their ionization  degree should allow 
detection  in the UV or in X-rays, as in ordinary AGNs. Since the wind
remains relatively cold on rather  small, sub-kpc scales, the actual 
detectability of
the wind may be complicated by  other  factors, like the  absorption by 
intervening  clouds. 
The slow outflow is active during a small fraction ($10-30\%$) of the time.
Hence, most clusters will not show any signature of the slow wind.


\section{Summary}
\label{s:summa}

The purpose of the current paper is to propose a plausible
new ingredient for the moderate cooling flow (CF) model.
We propose (speculate) that a large fraction, or even most,
of the mass that cools to low temperatures is ejected back to
the hot ICM via an accretion disk around the central black hole.
In the moderate CF model the effective age of
the CF is much shorter than the cluster age, because of
intermittent heating; most likely via AGN activity; only in
the very inner regions the flow is in a steady state phase;
cooling to low temperatures ($T< 10^4 \K$) occurs at a rate
much lower than in older versions of CF models (see \S~1).

This proposed scenario is based on several observations and theoretical
considerations (\S~3) that hint either at bubbles
temperatures of $T_b \lesssim 100 \keV$, or at the
possibility that AGN can blow relatively slow winds (jets),
$1000 \lesssim v_j \lesssim 10,000 \kms$.
With the lower energy per ejected unit mass (Equation~\eqref{e:temp}), the
outflowing mass  required to account for the energy in the bubbles
may comprise a significant fraction of the mass cooling
to low temperature (Equation~\eqref{e:mjet}).
In the proposed scenario, the average power of the AGN is much
higher than that inferred from radio emission, or even from
energy content of the large bubbles.
Some mass will be blown at $v_j \simeq 3000-5000 \kms$, forming
small bubbles, with density not much smaller than their surroundings.
\citet{Bir04} find that the energy associated with X-ray
deficient bubbles does not generally explain the low mass cooling
rate, unless the bubbles probe only a small fraction of the
total kinetic energy.
The proposed speculative scenario accounts for this
extra energy.

In the proposed scenario most of the cooling mass stays in
the inner region $r \lesssim 10-50 \kpc$ of the CF cluster,
as most bubbles do not buoy to large radii.
This is compatible with the implication that there is
no vigorous mixing in CF clusters \citep[][]{Boh04}.
We stress that there is no wind of the ICM, simply we proposed
that some fraction of the jets (or collimated outflow),
which inflate some of the bubbles, is blown at a relatively
low speed, and contains significant fraction of the mass
cooling to temperatures of $\lesssim 10^4 \K$.
This ingredient of the moderate CF model makes better the
agreement of the model with observations.
However, this ingredient is not necessary for the
moderate CF model.

The proposed scenario predicts the following.
\begin{enumerate}
\item As being an ingredient in the moderate CF model,
it requires that some mass cools to low temperatures,
but at a moderate average rates
$\dot M_{\rm cool} \simeq 1-50 \msyr$.
\item  The temperatures of the gas in large bubbles is, in most cases,
$T_b \lesssim 100 \keV$.
\item In many clusters small bubbles exist, which are
only slightly hotter than their surroundings.
Such bubbles may reveal themselves as patchy low-X-ray emitting
regions, such as those observed in M87-Virgo
\citep[see the X-ray image by][]{You02}.
\item 
Weak radio emission between X-ray ripples, as the ripples discovered by
\citet{Fab03c} in the Perseus cluster.  We recognize, however, 
that these observations may be difficult, chiefly
on account of confusing projection effects.
\item Sub-relativistic, $v_j \lesssim 10,000 \kms$, flow from
relatively extended region around the central black hole
of CF clusters exists from time to time 
(or presently, for $10-30\%$ of the cooling-flow clusters).
\item At least in some favourable cases, we do expect to  
detect such outflows thanks to their absorption in the UV or X-ray spectrum, 
as  observed in the  winds blown by  other AGNs.

\end{enumerate}

The first three predictions can be tested with present X-ray telescopes,
We predict that deep X-ray observations of the inner regions of CF
clusters will reveal:
(1) emission from gas cooling below $\sim 1 \keV$, as in Abell~2597
\citep[][]{Mor04} and Abell~2029 \citep[][]{Cla04};
(2) some X-ray emission from some bubbles, implying
their density is not much lower than their ambient density; and
(3) small bubbles scattered around, only slightly hotter than
their surroundings.

\acknowledgements

We thank Orkan M.~Umurhan for comments on the original manuscript,
and Shai Kaspi for useful discussion. We also thank an anonymous referee for
insightful comments.
This research  was partially supported by a grant from the
Israel Science Foundation.
F. Pizzolato was supported also by grant No. 2002111 from the United
States-Israel Binational Foundation, Jerusalem, Israel. He is currently 
supported by a Fine Fellowship.



\end{document}